\documentclass[twocolumn,preprintnumbers,pra,superscriptaddress,showkeys]{revtex4}%smaller,showpacs,

\usepackage{amsmath,amssymb,amsfonts,bbm,graphicx,float,times,psfrag}
\usepackage[pdftex]{color}
\usepackage{amsmath,bm}
\usepackage[colorlinks, linkcolor=blue, citecolor=blue,  breaklinks=true]{hyperref}
\usepackage{mathtools,amsfonts,mathptmx}
\usepackage[utf8]{inputenc}
\usepackage[T1]{fontenc}
\usepackage{xcolor}
\usepackage{braket}
\usepackage[titletoc,title]{appendix}
\usepackage[nameinlink,capitalize]{cleveref}

\begin{document}
\title{Low threshold quantum correlations via synthetic magnetism in Brillouin optomechanical system}

\author{D. R. K. Massembele}
\email{kenigoule.didier@gmail.com}
\affiliation{Department of Physics, Faculty of Science, 
University of Ngaoundere, P.O. Box 454, Ngaoundere, Cameroon}

\author{P. Djorwé}
\email{djorwepp@gmail.com}
\affiliation{Department of Physics, Faculty of Science, 
University of Ngaoundere, P.O. Box 454, Ngaoundere, Cameroon}
\affiliation{Stellenbosch Institute for Advanced Study (STIAS), Wallenberg Research Centre at Stellenbosch University, Stellenbosch 7600, South Africa}

\author{K. B. Emale}
\email{emale.kongkui@facsciences-uy1.cm}
\affiliation{Department of Physics, Faculty of Science, University of Yaounde I, P.O.Box 812, Yaounde, Cameroon}

\author{Jia-Xin Peng}
\email{18217696127@163.com}
\affiliation{School of Physics and Technology, Nantong University, Nantong, 226019, People’s Republic of China}

% \author{A.K. Sarma}
% \email{aksarma@iitg.ac.in}
% \affiliation{Department of Physics, Indian Institute of Technology Guwahati, Guwahati 781039, India}

\author{A.H. Abdel-Aty}
\email{amabdelaty@ub.edu.sa}
\affiliation{Department of Physics, College of Sciences, University of Bisha, Bisha 61922, Saudi Arabia}

\author{K.S. Nisar}
\email{n.sooppy@psau.edu.sa}
\affiliation{Department of Mathematics, College of Science and Humanities in Alkharj, Prince Sattam Bin Abdulaziz University, Alkharj 11942, Saudi Arabia}

\begin{abstract}
We propose a scheme to generate low driving threshold quantum correlations in Brillouin optomechanical system based on synthetic magnetism. Our proposal consists of a mechanical (acoustic) resonator coupled to two optical modes through the standard optomechanical radiation pressure (an electrostrictive force). The electrostrictive force that couples the acoustic mode to the optical ones striggers Backward Stimulated Brillouin Scattering (BSBS) process in the system. Moreover, the mechanical and acoustic resonators are mechanically coupled through the coupling rate $J_m$, which is $\theta$-phase modulated. Without a mechanical coupling, the generated quantum correlations require a strong driving field. By accounting phonon hopping coupling, the synthetic magnetism is induced and the quantum correlations are generated for low coupling strengths. The generated quantum correlations  display sudden death and revival phenonmena, and are robust against thermal noise. Our results suggest a way for low threshold quantum correlations generation, and are useful for  quantum communications, quantum sensors, and quantum computational tasks.
\end{abstract}

\maketitle
\date{\today}

\section{Introduction}\label{intro}

Owing to the ongoing progresses in quantum science, quantum correlations have become interesting resources for numerous quantum applications. Inportant research activities are carried out in order to engineer quantum correlations, i.e., quantum entanglement \cite{Riedinger_2018,Kotler_2021,Thomas_2020,Djor.2014,Tchodimou.2017}, quantum discord \cite{livier.2001,Giorda_2010,Adesso_2010}, and quantum steering \cite{Kogias_2015,chen.2012,Kogia.2015} for instance. These quantum features are useful for quantum tasks such as quantum communication \cite{Li_2017}, quantum information processing \cite{Wendin.2017,Slussarenko.2019,Meher.2022}, quantum sensing \cite{Xia_2023,Degen_2017}, and quantum computing \cite{Zhong2020,Arute.2019,Zidan2021} to mention only few. The aforementioned quantum correlations can be synthetize in diverse physical systems including microwaves circuits \cite{Salma2023}, plasmonic systems \cite{Hu2016}, and optomechanical structures \cite{Purdy2017,Bemani2019,Riedin.2016}. 

Optomechanical structures, allowing an electromagnetic field and mechanical center of mass to interact through optomechanical coupling, have been used to fooster interesting applications ranging from classical domain to quantum regime.  In the classical domain, behaviors such as collective phenonmena \cite{Djorwe2018,Colombano2019,Djorwe.2020,Li_2022,Houwe2023}, nonlinear dynamic \cite{Roque.2020,Xu_2024,Foulla2017,Djor.2022} and chaos \cite{Navarro.2017,Zhu_2023,Stella_2023} have been uncovered. These phenonmena find application in random number generation \cite{Madiot_2022}, communication schemes based-synchronization \cite{Rodrigues_2021,Berra.2023}, and encryption schemes based-chaos \cite{Wen2023,Zhang.2023}. In the quantum regime, a plethora of quantum phenonmena have been induced such as squeezed states \cite{Qin_2022,Banerjee_2023}, ground states cooling \cite{Clark_2017,Jiang_2021,Djor.2012}, and quantum teleportation \cite{Fiaschi2021} to name only few. In this work, we focus our attention on the generation of entanglement and Gaussian quantum discord in Brillouin optomechanical system. Our scheme is based on a synthetic magnetism control, which allows to enhance quantum correlations in our proposal. Similar technic has been recently used to tune optomechanically-induced transparency \cite{Lai2020}, for multimode optomechanical cooling \cite{Huang2022}, and to generate entanglement in optomechanics \cite{Lai_2022,Li_2023,Nori2022}. In these investigations, the observed physical phenonmena were enhanced through the breaking of the dark mode effect by tuning both the phonon hopping rate $J_m$ and its phase $\theta$. 

In this work, we propose to enhance quantum correlations by controlling the synthetic magnetism in a Brillouin optomechanical system. Our benckmark system consists of a mechanical (acoustic) resonator coupled to two optical modes through a radiation pressure (electrostrictive force). The mechanical and acoustic resonators are mechanically coupled through a phonon hopping rate $J_m$, which is $\theta$-phase modulated. By adiabatical elimination of one optical field, the four mode system were reduced to a three mode one, and owing to the large acoustic dissipation strength, we were able to generate optomechanical quantum entanglement. However, this entanglement required a large strength of both optomechanical and acoustic coupling rates. In order to synthetize optomechanical entanglement with less threshold power, i.e. less optomechanical/acoustic coupling, we proceeded through a tuning of the phonon hopping rate $J_m$ its and phase $\theta$ into our analysis. Our analysis has led to the following findings, i) the synthetic magnetism effect results to a loss threshold quantum correlations generation in our proposal; ii) the generated quantum correlations display oscillatory patterns leading to sudden death and revival phenonmena, and iii) the quantum correlations and their robustness against thermal noise are enhanced over the phonon hopping rate. These results suggest a path towards a loss threshold generation of quantum correlations in optomechanical plateforms, and they can be useful for quantum communications, quantum sensors, and quantum computational tasks.

The rest of our work is organized as follow. In \autoref{sec:model} we discuss our proposed model,and  provide its dynamical equations. The enhancement of the quantum correlations, i.e., optomechanical entanglement and Gaussian quantum discord, are presented in \autoref{sec:entag} and \autoref{sec:disc}, respectively. We conclude our investigation in \autoref{sec:Concl}.

\section{Model and dynamical equations} \label{sec:model}
Our benckmark system consists of a mechanical resonator $b_m$ and an acoustic mode $b_a$ which couple each to two optical modes $a_{j=1,2}$. The mechanical resonator couples to the optical modes through the standard radiation pressure coupling, while the acoustic mode couples to the same optical modes via an electrostrictive force that induces a Backward Stimulated Brillouin Scattering (BSBS) process in the system. Moreover, the mechanical and acoustic resonators are mechanically coupled through the phonon hopping term $J_m$ that is modulated via a phase $\theta$. Such a phase modulation results from a phase difference of individual driving phase, and striggers a synthetic magnetism into the system \cite{Schmidt.2015,Fang.2017,Brendel.2017}. The Hamiltonian capturing the dynamics of our system is ($\hbar=1$):
\begin{equation}\label{eq:eq1}
 H =H_{\rm{0}}+H_{\rm{OM}}+H_{\rm{BSBS}}+H_{\rm{drive}}, 
 \end{equation}
where
\begin{eqnarray}  \label{eq:eq2}
H_{\rm{0}}&:=&\sum_{j=1,2} \omega_{c_j} a_j^\dagger  a_j + \omega_m b_m^\dagger b_m + \omega_a b_a^\dagger b_a, \nonumber\\ 
H_{\rm{int}}&:=&\sum_{j=1,2}g_{m_j} a_j^\dagger a_j   (b_m +b_m^\dagger)+{J_m}({e^{i\theta}}{b_{a}^{\dagger}}{b_m} + {e^{-i\theta}}{b_a}{b_{m}^{\dagger}}), \nonumber\\
H_{\rm{BSBS}}&:=&- g_a( a_1^\dagger a_2 b_a+ a_1 a_2^\dagger b_a^\dagger),\nonumber\\
H_{\rm{drive}}&:=&\sum_{j=1,2}iE_j(a_j^{\dagger }e^{-i\omega_{p_j} t} - a_je^{i\omega_{p_j} t}).\nonumber
\end{eqnarray} 
The free Hamiltonian of our system is captured by $H_{\rm{0}}$ where the optical, the mechanical and the acoustic free energie are described by the first, second and the third term, respectively. The optomechanical interaction together with the mechanical hopping coupling term are captured by $H_{\rm{int}}$. The BSBS process and the drivings are described by the Hamiltonian $H_{\rm{BSBS}}$ and $H_{\rm{drive}}$, respectively. The optical modes are represented by their annihilation (creation) operators $a_j$ ($a_j^{\dagger}$), while the mechanical (acoustic) operator is $b_m$ ($b_a$). The other parameters of our system are the single-photon optomechanical couplings are $g_{m_j}$, the acoustic coupling $g_a$, the driving amplitude $E_j$, the driving frequency $\omega_{p_j}$, the optical cavity frequency $\omega_{c_j}$, and the mechanical (acoustic) frequency $\omega_m$ ($\omega_a$).  For now on, we assume $g_{m_j}\equiv g_m$.

\begin{figure}[tbh]
  \begin{center}
  \resizebox{0.4\textwidth}{!}{
  \includegraphics{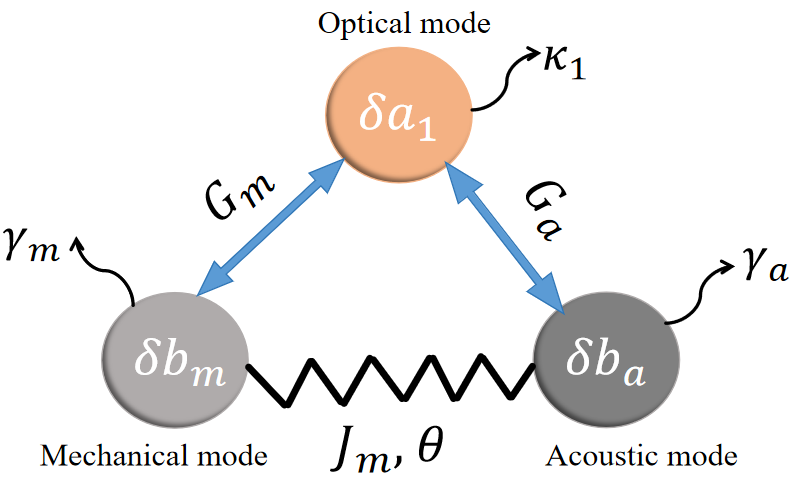}}
  \end{center}
\caption{Sketch of the linearized three modes optomechanical system. The  optical mode $\delta a_1$ is coupled to the acoustic (mechanical) mode $\delta b_a$ ($\delta b_m$) through the coupling $G_a$ ($G_m$), which are induced by electrostrictive (radiation pressure) force. The phonon-phonon hopping rate $J_m$ is modulated by the phase $\theta$. The dissipation of the mechanical, optical, and acoustic mode are $\gamma_m$, $\kappa_1$, and $\gamma_a$, respectively.}
\label{fig:Fig1}
\end{figure}

By moving into the frame rotating at $H_r=\omega_{p_1}a_1^\dagger a_1 +\omega_{p_2} a_2^\dagger a_2 +({\omega_{p_1}-\omega_{p_2}})b_a^\dagger b_a $, our Hamiltonian described in \autoref{eq:eq1} yields,
\begin{eqnarray}\label{eq:eq3}
H'&=- \Delta_1 a_1^\dagger  a_1 + \Delta_a b_a^\dagger b_a+\omega_m b_m^\dagger b_m- g_{m} a_1^\dagger a_1 (b_m +b_m^\dagger)\nonumber \\&+{J_m}({e^{i\theta}}{b_{a}^{\dagger}}{b_m} + {e^{-i\theta}}{b_a}{b_{m}^{\dagger}})+iE_1(a_1^{\dagger } - a_1) \nonumber \\&- G_a( a_1^\dagger b_a+ a_1 b_a^\dagger),
\end{eqnarray}
where the following detunings have been defined $\Delta_1=\omega_{p_1}-\omega_{c_1}$, and $\Delta_a=\omega_a+\omega_{p_2}-\omega_{p_1}$. Moreover, the effective acoustic coupling ${G_a}=g_a \alpha_2$ has been introduced, where $\alpha_2$ stands for the steady-state of the optical mode $a_2$. We have treated the control optical mode $a_2$ classically, since it is assumed to be strong compared to the weak strength of both the Brillouin acoustic mode $b_a$ and the optical mode $a_1$ (see  \cref{App} for details). Following the usual linearization process in optomechanics, where the operators fields are split into their mean values with some amount of fluctuation $\mathcal{O}=\langle\mathcal{O}\rangle+\delta \mathcal{O}$ ($\mathcal{O}\equiv a_1,b_a,b_m $), \autoref{eq:eq3} can be linearized, leading to the three modes optomechanical system sketched in \autoref{fig:Fig1}.
This linearization process allows to derive the following dynamical set of equations for the fluctuations (see \cref{App} for details),  
\begin{equation}\label{eq:fluc}
\begin{cases}
\delta\dot{a}_1 &= \left(i\tilde{\Delta} - \frac{\kappa}{2}\right) \delta a_1 + iG_c({\delta}b_{m}^{\dagger} +{\delta}b_m) +i{G_b} \delta b_a \\&+\sqrt{\kappa}a_1^{in}\\ 
\delta\dot{b}_a &= - ({\frac{\gamma_a}{2}} + i{\Delta_a})\delta{b}_a  - i{J_{m}}e^{i\theta}\delta{b}_m +i{G_a} \delta a_1 \\&+\sqrt{\gamma_a}b_{a}^{in} \\
\delta\dot{b}_m &= -({\frac{\gamma_m}{2}}+ i {\omega_m})\delta{b}_m   -i{J_{m}}e^{-i\theta}\delta{b}_a + i(G_m^*\delta a_1 \\&+ G_m\delta a_1^\dagger ) + \sqrt{\gamma_m}b_{m}^{in}, 
\end{cases}
\end{equation}
where the optical ($\kappa$), mechanical ($\gamma_m$) and acoustic ($\gamma_a$) dissipation have been introduced. The effective detuning $\tilde{\Delta}=\Delta-2g_m\rm Re(\beta_m)$ has been defined, with $\beta_m$ the steady-state of the mechanical mode. Moreover, the zero-mean noise operators $a_1^{in}$, $b_a^{in}$ and $b_m^{in}$ are introduced, which are characterized by the auto-correlation functions,
\begin{eqnarray}\label{eq:noise}
&\langle a_1^{in}(t)a_1^{in\dagger}(t') \rangle =\delta(t-t'), \hspace*{0.5cm} \langle a_1^{in\dagger}(t)a_1^{in}(t') \rangle = 0,\\
&\langle b_a^{in}(t)b_a^{in\dagger}(t') \rangle = \delta(t-t'), \hspace*{0.5cm} \langle b_a^{in\dagger}(t)b_a^{in}(t') \rangle = 0 ,\\
&\langle b_m^{in}(t)b_m^{in\dagger}(t') \rangle = (n_{th}^j+1)\delta(t-t'), \\ 
&\langle b_m^{in\dagger}(t)b_m^{in}(t') \rangle = n_{th}^j\delta(t-t'), \nonumber \\ 
\end{eqnarray}
where $n_{th}$ is the thermal phonon occupation of the mechanical resonator defined as $n_{th}=[\rm exp(\frac{\hbar \omega_m}{k_bT})-1]^{-1}$, with the Boltzmann constant $\rm k_b$. In this analysis, the thermal acoustic occupancy has been neglected owing to the high-frequency Brillouin mode $b_a$ ($\omega_m \ll \omega_a$). In order to investigate on the entanglement feature, we introduce the following quadrature operators, $\delta X_{\mathcal{O}} =\frac{\delta \mathcal{O}^\dagger + \delta \mathcal{O} }{\sqrt{2}}$, $\delta Y_{\mathcal{O}} =i\frac{\delta \mathcal{O}^\dagger - \delta \mathcal{O}}{\sqrt{2}}$, together with their related noise quadratures, $\delta X^{in}_{\mathcal{O}} =\frac{\delta \mathcal{O}^{\dagger in} + \delta \mathcal{O}^{in}}{\sqrt{2}}$, $\delta Y^{in}_{\mathcal{O}} =i\frac{\delta \mathcal{O}^{\dagger in} - \delta \mathcal{O}^{in}}{\sqrt{2}}$, where $\mathcal{O}\equiv a_1,b_a,b_m$. These quadratures lead to a new set of  dynamical equations written in its compact form, 
\begin{equation}\label{eq:quadra}
\dot{u}={\rm A} u+u^{in},
\end{equation}
where ${u}=(\delta X_{a_1},\delta Y_{a_1}, \delta X_{b_a},\delta Y_{b_a},\delta X_{b_m},\delta Y_{b_m})^T$, $u^{in}=(\sqrt{\kappa} \delta X_{a_1}^{in},\sqrt{\kappa} \delta Y_{a_1}^{in},\sqrt{\gamma_a} \delta X_{b_a}^{in},\sqrt{\gamma_a} \delta Y_{b_a}^{in},\sqrt{\gamma_m} \delta X_{b_m}^{in},\sqrt{\gamma_m} \delta Y_{b_m}^{in},)^T$ with the matrix $\rm A$,
\begin{equation}\label{eq:matrix}
{\rm A}=
\begin{pmatrix}
-\frac{\kappa_1}{2}&-\tilde{\Delta}&0&-G_a&0&0 \\
\tilde{\Delta}&-\frac{\kappa_1}{2}&G_a&0&2G_m&0 \\ 
0&-G_a&-\frac{\gamma_a}{2}&\Delta_a& 0 & 0 \\
G_a&0&-\Delta_a&-\frac{\gamma_a}{2}&0 & 0 \\
0&0&0&0&-\frac{\gamma_m}{2}&\omega_m \\
2G_m&0&0&0&-\omega_m& -\frac{\gamma_m}{2}
\end{pmatrix}.
\end{equation}
For simplicity in our analysis, we have assumed that the effective couplings $G_m$ and $G_a$ are real numbers.

\section{Enhancement of steady-state optomechanical entanglement through the synthetic magnetism}\label{sec:entag}
The steady-state optomechanical entanglement is analyzed by evaluating the covariance matrix whose elements are defined as $V_{ij}=\frac{\langle u_i u_j  + u_j u_i \rangle}{2}$, which also satisfy the motional equation,
\begin{equation}\label{eq:lyad}
\dot{V}={\rm A}V+V{\rm A^T}+D,
\end{equation}
where $D$ is the diagonal diffusion matrix expressed as $D=\rm{Diag}[\frac{\kappa}{2},  \frac{\kappa}{2}, \frac{\gamma_a}{2}, \frac{\gamma_a}{2}, \frac{\gamma_m}{2}(2n_{th} + 1), \frac{\gamma_m}{2}(2n_{th} + 1)]$. To carry out the entanglement analysis, the matrix $\rm A$ must satisfy the Routh-Huritz stability criterion i.e., all its eigenvalues should have negative real parts \cite{DeJesus}. With our used parameters, this condition has been fulfilled. Moreover, the steady-state entanglement is captured under the condition that the dynamical variables in \autoref{eq:lyad} are no longer time dependent, reducing this equation to the following Lyaponuv equation,

\begin{equation}\label{eq:lyap}
{\rm A}V+V{\rm A^T}=-D.
\end{equation}
The $V_{ij}$ elements of the covariance matrix can be computed numerically, and the matrix  $V$ can be written on its general form, 
\begin{equation}\label{cov}
\rm{V}=
\begin{pmatrix} 
V_{\alpha}&V_{\alpha,\beta}&V_{\alpha,\gamma} \\
V_{\alpha,\beta}^{\intercal}&V_{\beta}&V_{\beta,\gamma} \\ 
V_{\alpha,\gamma}^{\intercal}&V_{\beta,\gamma}^{\intercal}&V_{\gamma}
\end{pmatrix},
\end{equation}
where $V_i$ and  $V_{ij}$ represent blocs of $2\times2$ matrices (with $i,j \equiv \alpha,\beta,\gamma$). The diagonal blocs $V_i$ correspond to the optical mode ($i = \alpha$), the mechanical mode ($i = \beta$), and the acoustic mode ($i = \gamma$), respectively. The off-diagonal blocks capture the correlations between different subsystems. For instance, $V_{\alpha,\beta}$ describes the correlations between the driving field and the mechanical resonator, $V_{\alpha,\gamma}$ describes the correlations between the driving field and the acoustic mode, while $V_{\beta,\gamma}$ stands for the correlations between the mechanical and the acoustic mode. The bipartite entanglement within two subsystems is then quantified by the logarithmic negativity ($E_N$), which is evaluated by tracing out the non-necessary third mode. This logarithmic negativity $E_N$ is defined as,
\begin{equation}\label{eq:en}
E_N=\rm max[0,-\ln(2\nu^-)], 
\end{equation}
where $\nu^- = 2^{-1/2}\left[ \Delta_{\chi}-\sqrt{\Delta_{\chi}^2-4I_4}\right]^{1/2}$. There is an entanglement in the system when the condition $\nu^-<1/2$ is fulfilled, which is equivalent to Simon's necessary and sufficient entanglement criterion for Gaussian states. The covariance matrix $\chi$, for the targeted subsystems can be defined as,
\begin{equation}\label{eq:QD}
\rm{\chi}=\begin{pmatrix} 
V_i&V_{ij} \\
V_{ij}^{\intercal}&V_j
\end{pmatrix},
\end{equation}
so that $\Delta_{\chi}=\rm{I_1+I_2-2I_3}$, where we have defined four symplectic invariants as $I_1=\rm det V_i$, $I_2=\rm det V_j$, $I_3=\rm det V_{ij}$, and $I_4=\rm det \chi$. From now on, we will consider that our system dwells into the red-sideband regime for the mechanical resonator ($\tilde{\Delta}=-\omega_m$), which is a prerequisite for cooling that is a requirement for entanglement engineering.
\begin{figure}[tbh]
\begin{center}
  \resizebox{0.5\textwidth}{!}{
  \includegraphics{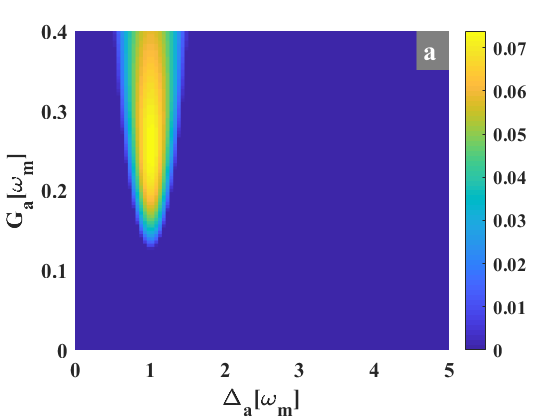}}
  \resizebox{0.5\textwidth}{!}{
  \includegraphics{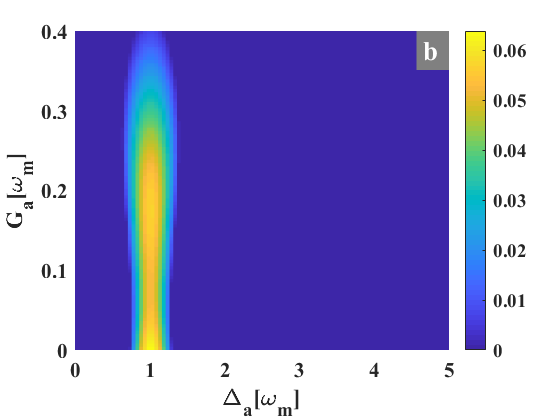}}
  \end{center}
\caption{Contour of the optomechanical entanglement versus $G_a$ and $\Delta_a$ for $J_m=0$ and $J_m=0.2\omega_m$ (with $\theta=\frac{\pi}{2}$) in (a) and (b), respectively. The parameters used are $\omega_m/2\pi=1\rm {MHz}$, $g_m=10^{-4}\omega_m$, $\kappa=0.02\omega_m$, $\gamma_a=0.4\omega_m$, $\gamma_m=10^{-4}\omega_m$, $n_{th}=100$, $G_m=0.15\omega_m$, and $\tilde{\Delta}=-\omega_m$.}
\label{fig:Fig2}
\end{figure}

Regarding the sideband of the acoustic mode, we need to look for optimal acoustic parameters that lead to large entanglement generation. Among these parameters we have the effective coupling ($G_a$), and the effective detuning ($\Delta_a$). To figure out our entanglement analysis, we use the following state-of-the-art optomechanical parameters, i.e., $\omega_m/2\pi=1\rm {MHz}$, $g_m=10^{-4}\omega_m$, $\kappa=0.02\omega_m$, $\gamma_m=10^{-4}\omega_m$, $n_{th}=100$, $G_m=0.15\omega_m$, and $\tilde{\Delta}=-\omega_m$. For the acoustic dissipation,  we choose it larger than both $\kappa_1$ and $\gamma_m$ as usually assumed in BSBS system. Therefore, have set $\gamma_a=0.4\omega_m$, and this large value of the acoustic decay rate will be confirmed later on through our results.  To get the optimal values of the acoustic parameters, we represent in \autoref{fig:Fig2} the contour plot of the optomechanical entanglement versus the effective coupling $G_a$ and the detuning $\Delta_a$. In \autoref{fig:Fig2}a, the phonon hopping rate is not accounted ($J_m=0$), while $J_m=0.2\omega_m$ for \autoref{fig:Fig2}b. It can be seen that large entanglement is generated for the acoustic detuning centered around $\Delta_a=\omega_m$. For the effective coupling $G_a$, there is a certain threshold from where the entanglement is generated in our system (for $J_m=0$) as shown in \autoref{fig:Fig2}a. This threshold value which is around $G_a\sim0.12\omega_m$, reveals how the BSBS process is a key ingredient for the engineering of optomechanical entanglement in our proposal. By taking into account (for $J_m\neq0$), \autoref{fig:Fig2}b shows that there is no threshold value of $G_a$ for the entanglement generation around $\Delta_a=\omega_m$. This means that the synthetic magnetism enhances entanglement in our system and it even extends the entanglement generation into the regime where the BSBS effect is weak ($G_a\sim0$). From now on, we will use the optimal detuning $\Delta_a=\omega_m$, and the coupling $G_a$ will be indicated based on what is depicted in \autoref{fig:Fig2}.      

\begin{figure*}[tbh]
\begin{center}
  \resizebox{0.45\textwidth}{!}{
  \includegraphics{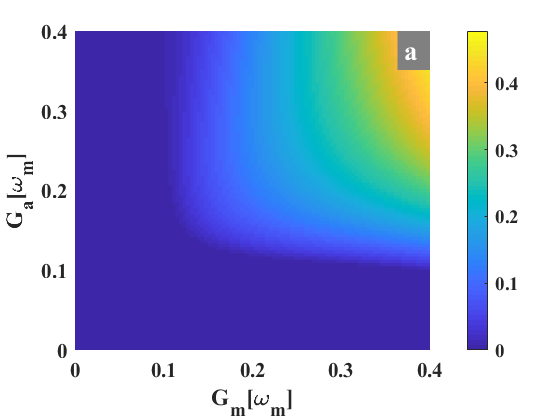}}
  \resizebox{0.45\textwidth}{!}{
  \includegraphics{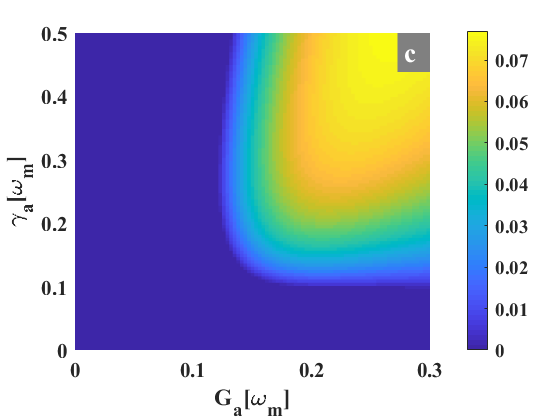}}
  \resizebox{0.45\textwidth}{!}{
  \includegraphics{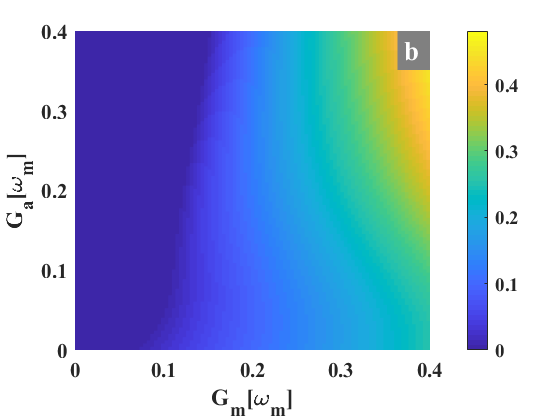}}
  \resizebox{0.45\textwidth}{!}{
  \includegraphics{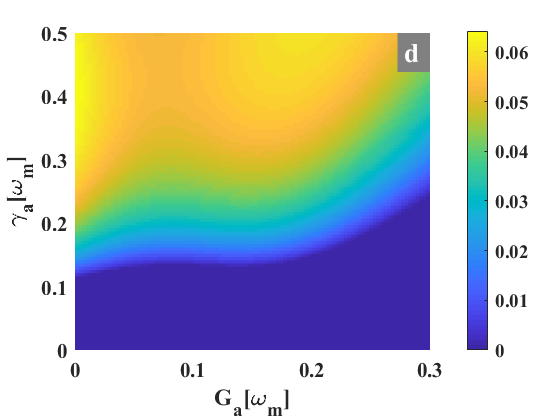}}
  \end{center}
\caption{Contour plot of optomechanical entanglement versus $\gamma_a$ and $G_a$ for $J_m=0$ and $J_m=0.2\omega_m$ (with $\theta=\frac{\pi}{2}$) in (a) and (b), respectively. Contour plot of the optomechanical entanglement versus $\gamma_a$ and $G_a$ for $J_m=0$ and $J_m=0.2\omega_m$ (with $\theta=\frac{\pi}{2}$) in (c) and (d), respectively. In (c) and (d), we have set $G_m=0.15\omega_m$, and the phonon number is $n_{th}=100$ for all plots. The other parameters are the same as those in \autoref{fig:Fig2}.}
\label{fig:Fig3}
\end{figure*}

As the optomechanical entanglement depends also on the optomechanical coupling strength, we display in \autoref{fig:Fig3} (a,b) the contour plot of the entanglement versus the couplings $G_a$ and $G_m$. When $J_m=0$, \autoref{fig:Fig3}a shows how both couplings  $G_a$ and $G_m$ require a similar threshold from where entanglement is generated in our system. Interestingly, our fixed value of $G_m=0.15\omega_m$ is well into the admitted range values allowing entanglement generation as shown in \autoref{fig:Fig3}a. For a non-zero value of the phonon coupling ($J_m=0.2\omega_m$), it can be seen on \autoref{fig:Fig3}b that the acoustic coupling $G_a$ induces entanglement even for weak effect of BSBS as shown in \autoref{fig:Fig2}b. For the optomechanical coupling, however, it still requires a threshold value around $G_m=0.1\omega_m$ for the entanglement generation, which is lower than for the case $J_m=0$ (see \autoref{fig:Fig3}a).  From this analysis, it appears that the synthetic magnetism mosthly assists the BSBS effect than the optomechanical one. It results that the synthetic magnetism induces low threshold driving strength for the entanglement generation. In order to confirm the large values of the acoustic decay rate assumed in our analysis, we have display in \autoref{fig:Fig3}(c,d) the contour plot of the entanglement versus the $\gamma_a$ and $G_a$. When $J_m=0$, it can be seen that the acoustic decay rate has a threshold value around $\gamma_a=0.1\omega_m$ from where entanglement is striggered into the system  (see \autoref{fig:Fig3}c). By taking into account the synthetic magnetism, the acoustic dissipation still remains large while the effective coupling $G_a$ is extended to weak values as aforementionned in \autoref{fig:Fig2}b and \autoref{fig:Fig3}b. From \autoref{fig:Fig3} (c,d), it appears that the synthetic magnetism does not has an impact on the acoustic dissipation regarding the entanglement generation. Moreover, this dissipation remains large compared to the optical ($\kappa_1=2\times10^{-2}\omega_m$) and mechanical dissipation ($\gamma_m=10^{-4}\omega_m$) as assumed throughout our numerical analysis.        

\begin{figure}[tbh]
\begin{center}
  \resizebox{0.4\textwidth}{!}{
  \includegraphics{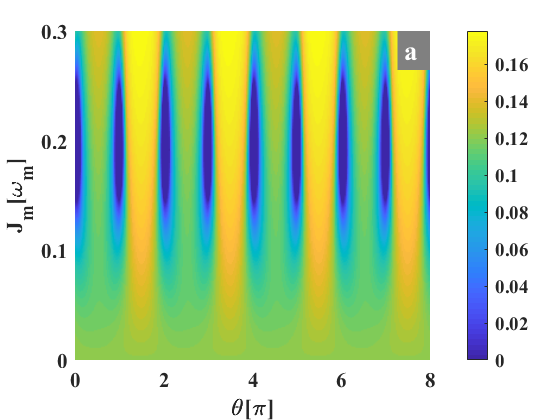}}
  \resizebox{0.4\textwidth}{!}{
  \includegraphics{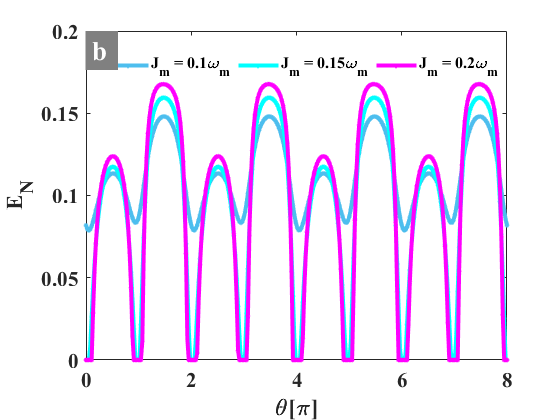}}
  \end{center}
\caption{(a) Contour plot of the optomechanical entanglement $E_N$  versus the phonon hopping rate $J_m$ and the modulation phase $\theta$. (b) $E_N$ versus $\theta$ for different values of $J_m$, i.e., $J_m=0.1\omega_m$  for the full line, at $J_m=0.15\omega_m$  for the dashed line, and at $J_m=0.2\omega_m$ for the dash-dotted line. For these figures we have set $G_a=0.2\omega_m$, $G_m=0.2\omega_m$, $n_{th}=100$, and the rest of the parameters are the same as those in \autoref{fig:Fig2}.}
\label{fig:Fig4}
\end{figure}

In order to get insight into the enhancement of the optomechanical entanglement under the synthetic magnetism, we represent \autoref{fig:Fig4}a that displays the contour plot of the entanglement versus the phonon hopping rate $J_m$ and its phase modulation $\theta$. This figure shows how the entanglement strength grows as the phonon hopping rate increases. Moreover, this enhancement of the optomechanical strength happens at specific values of the phase $\theta$, i.e., at $\theta= (n+\frac{1}{2})\pi$ for $n\in \mathbb{N}$. More interstingly, it can be seen that this enhancement is not symmetric around these particular values of $\theta$ depending on $n$. Indeed, one observes on \autoref{fig:Fig4}a that the strong enhancement happens for odd values of $n$ compared to the even values of $n$. It is noteworthy to mention that there is no (or less) entanglement for the values of $\theta= n\pi$ for $n\in \mathbb{N}$. Furthermore, the effect of $\theta$ is appreciated for weak values of the phonon hopping rate $J_m$. As the strength of $J_m$ increases, the generated optomechanical entanglement striggers nice oscillatory patterns along the phase axis, and these patterns display perfect oscillations around $J_m=0.2\omega_m$. Such oscillatory behaviors are depicted on \autoref{fig:Fig4}b, where the entanglement is plotted over the phase modulation $\theta$ for different values of $J_m$. As aforementioned in \autoref{fig:Fig4}a, it can be observed in \autoref{fig:Fig4}b how the entanglement gets stronger as the phonon hopping rate $J_m$ increases, and the three different behaviors related to the phase $\theta$ can be also seen. In particular for $J_m=0.2\omega_m$, the entanglement is not only highly enhanced, but it also reaches zero for $\theta= n\pi$. This specific behavior where the entanglement is lost during the oscillatory cycles is reminiscent of the entanglement death and reveal \cite{Chakrabo.2019}. Such a behavior is interesting for entanglement engineering since it allows to generate entanglement on demand by tuning specific  system parameters. The above analysis reveals that the synthetic magnetism is a tool to synthetize entanglement by tuning both the phonon rate $J_m$ and the phase $\theta$. Such generated entanglement may find applications in quantum information processing,  quantum sensing, and other quantum technologies. Moreover, the scheme presented here can be extended to other similar systems.        

\begin{figure}[tbh]
\begin{center}
  \resizebox{0.4\textwidth}{!}{
  \includegraphics{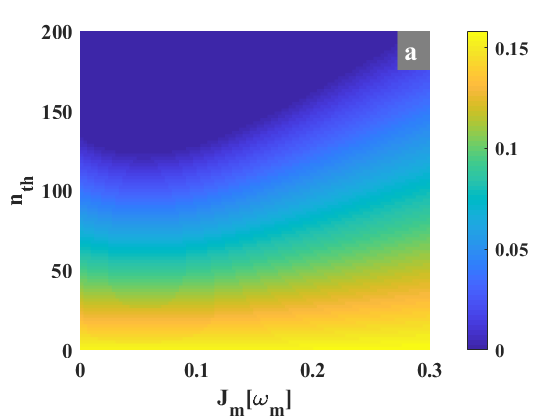}}
  \resizebox{0.4\textwidth}{!}{
  \includegraphics{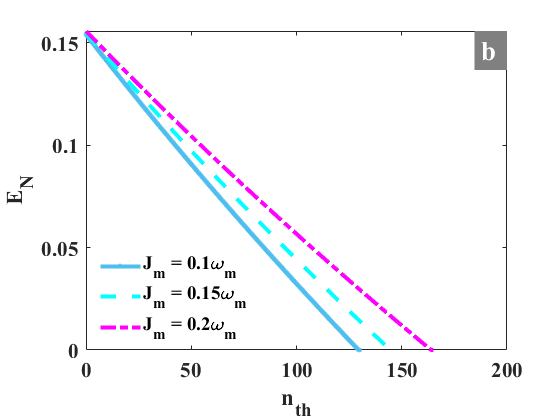}}
  \end{center}
\caption{(a) Contour plot of the optomechanical entanglement  versus the thermal phonon number $n_{th}$ and the phonon hopping rate $J_m$. (b) $E_N$ versus  $n_{th}$ for different values of $J_m$, i.e., $J_m=0.1\omega_m$  for the full line, at $J_m=0.15\omega_m$  for the dashed line, and at $J_m=0.2\omega_m$ for the dash-dotted line. For these figures we have fixed $G_a=G_m=0.15\omega_m$, $\theta=\frac{\pi}{2}$, and the rest of the parameters are the same as those in \autoref{fig:Fig2}.}
\label{fig:Fig5}
\end{figure}

Another interesting analysis is the robustness of the entanglement against thermal noise. Such an analysis reveals how strong the generated entanglement persists when the system is in contact with its surronding thermal bath. For this purpose, we investigated in \autoref{fig:Fig5} the influence of the thermal phonon number on the entanglement. For instance, \autoref{fig:Fig5}a displays the contour plot of the entanglement versus the thermal phonon number $n_{th}$ and the phonon hopping rate $J_m$. It can be seen that the entanglement becomes weak as the thermal phonon number increases. Moreover, one also observes that this entanglement resists better to the thermal noise as the phonon hopping rate $J_m$ grows. This behavior is highlighted in \autoref{fig:Fig5}b, where we have represented the entanglement versus the thermal phonon number for different values of $J_m$. This figure shows how for $J_m=0.1\omega_m$, the entanglement disappears before $n_{th}=150$. For $J_m=0.2\omega_m$, however, the entanglement persists well beyond $n_{th}=150$. This behavior reveals how the phonon hopping rate not only enhances entanglement in our proposal, but it also improves the robustness of the entanglement against the thermal noise.

\section{Enhancement of the quantum discord via the synthetic magnetism}\label{sec:disc}
Another interesting quantity to evaluate in our system is the quantum discord ($\epsilon_{QD}$), which captures the quantumness of the correlations in the state of a quantum system, even for separable states. Indeed, separability has often been thought as a synonymous of classicality, which is however not always the case \cite{livier.2001}. In that sense, quantum discord describes quantumness of correlations  beyond what does entanglement, since $\epsilon_{QD}$ cannot be captured by addressing entanglement only. Therefore, quantum discord is a more fundamental and interesting resource for quantum information processing tasks, and other quantum technologies. For a given quantum system, a measure of quantum discord that falls between zero and one ($0\leq\epsilon_{QD}\leq1$) means that the states of the system are either separable or entangled, whereas they are all entangled for $\epsilon_{QD}>1$ \cite{Giorda_2010}. For a given system described by the matrix as in \autoref{eq:QD}, the related Gaussian quantum discord is quantified as,
\begin{equation}
\epsilon_{QD}=f(\sqrt{I_2})-f(\eta^-)-f(\eta^+)+f(\sqrt{\epsilon}),
\end{equation}
where the function $f$ is defined by, 
\begin{equation}
f(x)=(x+\frac{1}{2})\ln(x+\frac{1}{2})-(x-\frac{1}{2})\ln(x-\frac{1}{2}).
\end{equation}
The two involved simplectic eigenvalues $\eta^-$ and $\eta^+$ are, 
\begin{equation}
\eta^{\pm}\equiv2^{-1/2}\left[\tilde\Delta_{\chi}\pm \sqrt{\tilde\Delta_{\chi}^2-4I_4}\right]^{1/2},
\end{equation} 
with $\tilde\Delta_{\chi}=\rm{I_1+I_2+2I_3}$, and $\epsilon$ reads,
\begin{equation}
\epsilon=
\begin{cases}
\left(\frac{2|I_{3}|+\sqrt{4I_{3}^{2}+(4I_{1}-1)(4I_{4}-I_{2})}}{(4I_{1}-1)}\right)^2~~\text{if}~~ \frac{4(I_{1}I_{2}-I_{4})^2}{(I_{2}+4I_{4})(1+4I_{1})I^{2}_{3}}\leq 1,\\
\frac{I_{1}I_{2}+I_4-I^{2}_3-\sqrt{(I_{1}I_{2}+I_{4}-I^{2}_{3})^2-4I_{1}I_{2}I_{4}}}{2I_{1}}~~\text{otherwise}.
\end{cases}
\end{equation}

\begin{figure}[tbh]
\begin{center}
  \resizebox{0.45\textwidth}{!}{
  \includegraphics{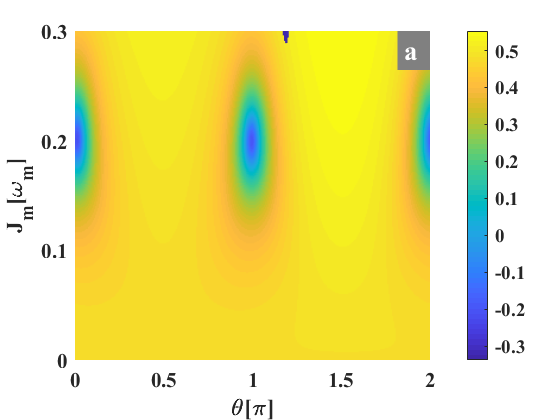}}
  \resizebox{0.45\textwidth}{!}{
  \includegraphics{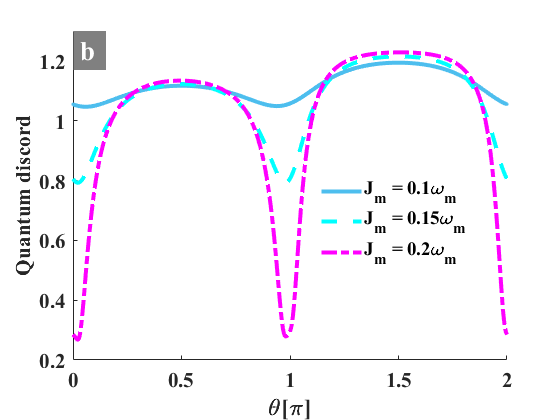}}
  \end{center}
\caption{(a) Contour plot of Quantum Discord ($\epsilon_{QD}$) versus the phonon hopping rate $J_m$ and the modulation phase $\theta$. (b) Extracted $\epsilon_{QD}$ from (a) for different values of $J_m$, i.e., $J_m=0.1\omega_m$  for the full line, at $J_m=0.15\omega_m$  for the dashed line, and at $J_m=0.2\omega_m$ for the dash-dotted line. In these plots, $G_a=0.2\omega_m$ and the other parameters are the same as those in \autoref{fig:Fig2}.}
\label{fig:Fig6}
\end{figure}

\begin{figure}[tbh]
\begin{center}
  \resizebox{0.45\textwidth}{!}{
  \includegraphics{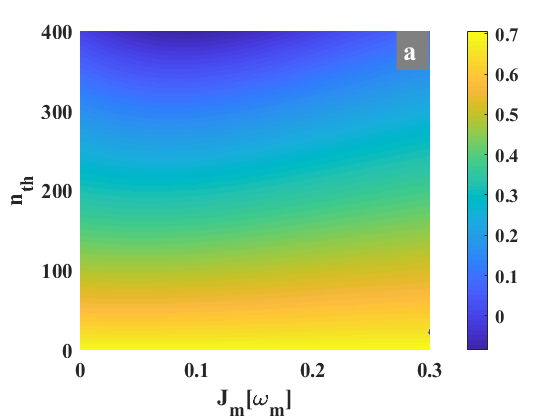}}
  \resizebox{0.45\textwidth}{!}{
  \includegraphics{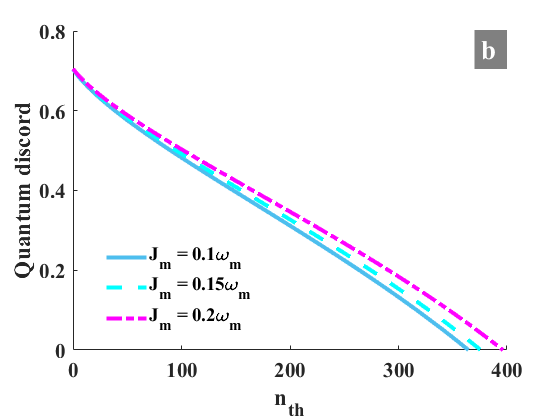}}
  \end{center}
\caption{Quantum discord behavior against thermal phonon number $n_{th}$. (a) Contour plot of $\epsilon_{QD}$ versus the thermal phonon number $n_{th}$ and the phonon hopping rate $J_m$. (b) Quantum discord versus the thermal phonon number for different values of phonon hopping rate. We have fixed $G_a=0.2\omega_m$ in (a), and the other parameters are the same as those in \autoref{fig:Fig2}.}
\label{fig:Fig7}
\end{figure}

\autoref{fig:Fig6} depicts the effect of the phonon hopping rate $J_m$ and its phase modulation $\theta$ on the Guassian quantity discord. Likewise as for the entanglement, It can be seen that $\epsilon_{QD}$ is improved for specific values of the phase, i.e., for $\theta=n\pi/2$ with $n$ being an odd interger (see \autoref{fig:Fig6}a). Such patterns displayed in \autoref{fig:Fig6}a reveals an oscillatory behavior of $\epsilon_{QD}$ along the phase as it can be seen in  \autoref{fig:Fig6}b. As the coupling strength $J_m$ is increasing, the  pattern oscillations of the quantum discord become more appreciated and the peak intensity of $\epsilon_{QD}$ slightly enhances around $\theta=n\pi/2$. Furthermore, \autoref{fig:Fig6}b exhibits a sudden death and revival behavior of the quantum discord. This feature reveals how the quantum discord can be engineered by tuning the phase $\theta$. With the values of $J_m$  used in \autoref{fig:Fig6}b, one observes that $\epsilon_{QD}>1$ around $\theta=n\pi/2$, ensuring entanglement as it can be seen from \autoref{fig:Fig4}b. However, one has $\epsilon_{QD}\approx0.24$ around $\theta=n\pi$ ($n$ being an integer) in \autoref{fig:Fig6}b, which leads to either an entangled or separable states as it can be seen in \autoref{fig:Fig4}b ($E_N\sim0$). Another interesting feature to analyze is the effect of thermal noise on quantum discord. Such an analysis is displayed by \autoref{fig:Fig7}. It can be seen that the quantum discord survives to thermal noise better than what entanglement does. Indeed, by comparing \autoref{fig:Fig5}a and \autoref{fig:Fig7}a, it appears that entanglement disappears for a phonon number $n_{th}=200$, while the quantum discord persists for phonon number up to $n_{th}\sim200$. Moreover, it can be also observed that the mechanical coupling slightly enhances the robustness of quantum discord against thermal noise (see \autoref{fig:Fig7}b). It results that quantum discord can be engineered in an environment where entanglement does not survives, revealing that quantum discord is an abundant quantum resource over entanglement. It results that quantum discord extends quantum applications beyond the regimes where entanglement is limited.    

\section{Conclusion}\label{sec:Concl}
This work investigated on low threshold quantum correlations generation through synthetic magnetism in a Brillouin optomechanical system. Our proposal consists of a mechanical (an acoustic) resonator which couples to two optical modes through the standard radiation pressure coupling (an electrostrictive force). Moreover, the mechanical and acoustic resonators are mechanically coupled through the phonon hopping term $J_m$ that is modulated via a phase $\theta$. Both the coupling terms $J_m$ and its related phase are used to induce low driving threshold quantum correlations. Without the synthetic magnetism, the generated quantum correlations require a strong driving field that correspond to large optomechanical/acoustic coupling. By introducing the synthetic magnetism, we greatly enhance quantum correlations in our system. Moreover, both entanglement and quantum discord display oscillatory patterns which are reminiscent of a sudden death and revival behavior of quantum correlations. It results that these quantum correlations can be engineered by tuning the phase modulation $\theta$. Furthermore, we have shown that these correlations are robust enough against thermal noise, since they persist for large values of thermal phonon number. However, it has been shown that quantum discord resists better to thermal noise than entanglement. This suggests that quantum discord extends quantum applications beyond what is allowed by entanglement, breaking down limitations of some quantum technologies.

\section*{Acknowledgments}
\textbf{Funding:} This work has been carried out under the Iso-Lomso Fellowship at Stellenbosch Institute for Advanced Study (STIAS), Wallenberg Research Centre at Stellenbosch University, Stellenbosch 7600, South Africa. K.S. Nisar is grateful to the funding from Prince Sattam bin Abdulaziz University, Saudi Arabia project number (PSAU/2024/R/1445). The authors are thankful to the Deanship of Graduate Studies and Scientific Research at University of Bisha for supporting this work through the Fast-Track Research Support Program.

% \textbf{Author Contributions:} P.D. and M.A. conceptualized the
% work and carry out the simulations and analysis. D.D., and B.D.-R. participated in all the discussions and provided useful suggestions to the final version of the manuscript. Y.P, D.D., and B.D.-R. supervised the work. All authors participated equally in the discussions and the preparation of the final manuscript.

%\textbf{Competing Interests:} All authors declare no competing interests.

% \section*{Data Availability}
% Relevant data are included in the manuscript
% and supporting information. Supplement data are available upon reasonable request.

\appendix \label{App} 

\section{Effective Hamiltonian}\label{App}
The Hamiltonian of our proposal in the natural frame is given by,
\begin{align}
H&=\sum_{j=1,2} \omega_{c_j} a_j^\dagger  a_j + \omega_a b_a^\dagger b_a + \omega_m b_m^\dagger b_m - \sum_{j=1,2}g_{m}a_j^\dagger a_j   (b_m +b_m^\dagger)\nonumber \\&+J_m({e^{i\theta}}{b_{a}^{\dagger}}{b_m} + {e^{-i\theta}}{b_a}{b_{m}^{\dagger}})+\sum_{j=1,2}iE_j(a_j^{\dagger }e^{-i\omega_{p_j} t} - a_je^{i\omega_{p_j} t})\nonumber \\&- g_a( a_1^\dagger a_2 b_a+ a_1 a_2^\dagger b_a^\dagger).
\end{align}
By moving to the frame rotating with the frequency $\omega_{p_1}a_1^\dagger a_1 +\omega_{p_2} a_2^\dagger a_2 +({\omega_{p_1}-\omega_{p_2}})b_a^\dagger b_a $, the above Hamiltonian becomes,

\begin{align}\label{eq:hamil}
H'&=-\sum_{j=1,2} \Delta_{j} a_j^\dagger  a_j +  \Delta_a b_a^\dagger b_a+\omega_m b_m^\dagger b_m- \sum_{j=1,2}g_{m}a_j^\dagger a_j (b_m +b_m^\dagger)\nonumber \\&+J_m({e^{i\theta}}{b_{a}^{\dagger}}{b_m} + {e^{-i\theta}}{b_a}{b_{m}^{\dagger}})+iE_1(a_1^{\dagger } - a_1) +iE_2(a_2^{\dagger} - a_2)\nonumber \\& - g_a( a_1^\dagger a_2 b_a+ a_1 a_2^\dagger b_a^\dagger),
\end{align}
with $\Delta_j=\omega_{p_j}-\omega_{c_j}$, and $\Delta_a=\omega_a+\omega_{p_2}-\omega_{p_1}$. 
By considering that the control field $a_2$ is strong enough compared to $a_1$, it can be treated classically by deriving its steady-state as,
\begin{align}
\dot{a}_2 &= i[H',a_2], \\
 &=(i\Delta_2-\frac{\kappa_2}{2})a_2 + ig_{m}a_2(b_m+b_m^\dagger) + E_2 + ig_a a_1 b_a^\dagger \nonumber \\
 &=(i\Delta'_2-\frac{\kappa_2}{2})a_2 + E_2 + ig_a a_1 b_a^\dagger,
\end{align}
with $\Delta'_2=\Delta_2 + g_{m}a_2(b_m+b_m^\dagger)$. The steady-state solution ($\dot{a}_2=0$) yields, 
\begin{align}
\alpha_{2}&\sim\frac{-E_2}{i\Delta'_2-\frac{\kappa_2}{2}} \hspace{1em} \text{or} \hspace{1em} |\alpha_{2}|\sim\frac{E_2}{\sqrt{\Delta_2^{'^2}+\frac{\kappa_2^2}{4}}}.
\end{align}
By replacing this expression in the rest of the Hamiltonian in \autoref{eq:hamil}, we get the following reduced Hamiltonian,
\begin{align}
H'&=-\Delta_{1} a_1^\dagger  a_1 +  \Delta_a b_a^\dagger b_a+\omega_m b_m^\dagger b_m- g_{m} a_1^\dagger a_1 (b_m +b_m^\dagger)\nonumber \\ &+J_m({e^{i\theta}}{b_{a}^{\dagger}}{b_m} + {e^{-i\theta}}{b_a}{b_{m}^{\dagger}})+iE_1(a_1^{\dagger } - a_1)- {G_a} ( a_1^\dagger b_a + a_1 b_a^\dagger ),
\end{align}
where the acoustic effective coupling is $G_a=g_a\alpha_{2}$ as mentionned in the main text.

By following the standard linearization process, where the operators fields are split into their mean values with some amount of fluctuation $\mathcal{O}=\langle\mathcal{O}\rangle+\delta \mathcal{O}$ ($\mathcal{O}\equiv a_1,b_a,b_m $), one gets the following mean dynamical set of equations,
\begin{align}\label{eq:meann}
\begin{cases}
\dot{\alpha_1}&=(i\tilde{\Delta}_1 -\frac{\kappa_1}{2}) \alpha_1 + iG_a \beta_a + E_1 \\
\dot{\beta_a}&=-(\frac{\gamma_a}{2}+i\Delta_a)\beta_a - iJ_m e^{i\theta}\beta_m + iG_a \alpha_1 \\
\dot{\beta_m}&=-(\frac{\gamma_m}{2}+i\omega_m)\beta_m - iJ_m e^{-i\theta}\beta_a + ig_{m_1}|\alpha_1|^2
\end{cases},
\end{align}
together with the dynamical fluctuation equations,
\begin{equation}\label{eq:flucc}
\begin{cases}
\delta\dot{a}_1 &= \left(i\tilde{\Delta}_1 - \frac{\kappa_1}{2}\right) \delta a_1 + iG_m({\delta}b_{m}^{\dagger} +{\delta}b_m) +i{G_a} \delta b_a \\&+\sqrt{\kappa}a_1^{in}\\ 
\delta\dot{b}_a &= - ({\frac{\gamma_a}{2}} + i{\Delta_a})\delta{b}_a  - i{J_{m}}e^{i\theta}\delta{b}_m +i{G_a} \delta a_1 \\&+\sqrt{\gamma_a}b_{a}^{in} \\
\delta\dot{b}_m &= -({\frac{\gamma_m}{2}}+ i {\omega_m})\delta{b}_m   -i{J_{m}}e^{-i\theta}\delta{b}_a + i(G_m^{\ast}\delta a_1 \\&+ G_m\delta a_1^\dagger ) + \sqrt{\gamma_m}b_{m}^{{in}}, 
\end{cases},
\end{equation}
where $\tilde{\Delta}_1=\Delta_1-2g_m \rm Re(\beta_m)$, and $G_m=g_{m_1}\alpha_1$.
The linearized Hamiltonian corresponding to \autoref{eq:flucc} yields,
\begin{align}
H_{lin}&=-\tilde{\Delta}_1 \delta a_1^\dagger  \delta a_1 +  \Delta_a \delta b_a^\dagger \delta b_a + (\omega_m +\Lambda)\delta b_m^\dagger \delta b_m- (G_m \delta a_1^\dagger \nonumber \\&+G_m^* \delta a_1) (\delta b_m +\delta b_m^\dagger)+{J_m}({e^{i\theta}}{\delta b_{a}^{\dagger}}{\delta b_m} + {e^{-i\theta}}{\delta b_a}{\delta b_{m}^{\dagger}}) \nonumber \\ & - G_a ( \delta a_1^\dagger \delta b_a + \delta a_1 \delta b_a^\dagger) .
\end{align}

\newpage

\bibliography{Squeezing}

\end{document}